\documentclass[doublecol,a4paper]{epl2} 
\usepackage{latexsym}
\usepackage{graphics}
\usepackage{epsfig}
\usepackage{amssymb}
\usepackage{amsmath}

\begin{document}
\title{Mass media and repulsive interactions in continuous-opinion dynamics}
\author{T. Vaz Martins\inst{1,2} \and M. Pineda\inst{3} \and R. Toral\inst{1}}
\institute{\inst{1}IFISC (Instituto de F\'isica Interdisciplinar y Sistemas Complejos), CSIC-UIB, Campus Universitat de les Illes Balears, E-07122 Palma de Mallorca, Spain, \\
\inst{2}Centro de 
F\'isica do Porto, Departamento de F\'isica e Astronomia, Faculdade de Ci\^encias da Universidade do Porto, 4169-007 Porto, Portugal, \\
\inst{3}Center for Nonlinear Phenomena and Complex Systems, Universit\'e Libre de Bruxelles, Code Postal 231, Campus Plaine, B-1050 Brussels, Belgium} 
\date{\today}


\abstract{
This letter focus on the effect of repulsive interactions on the adoption of an external message in an opinion model. With a simple change in the rules, we modify the Deffuant \emph{et al.} model to incorporate the presence of repulsive interactions. We will show that information receptiveness is optimal for an intermediate fraction of repulsive links. Using the master equation as well as Monte Carlo simulations of the message-free model, we identify the point where the system becomes optimally permeable to external influence with an order-disorder transition.
}
\maketitle
\section{Introduction}
\label{intro}

In recent years a lot of effort has been devoted to the study of opinion formation models using techniques borrowed from nonlinear and statistical physics \cite{castellano}. Models can be grouped in two big families, according to whether the variable that represents the opinion that an individual (agent) holds on a particular topic can take only a finite set of values or it is a continuous, real, variable. Among the latter, the celebrated Deffuant \emph{et al.} model \cite{def} allows opinions to evolve by means of a negotiation rule. A distinctive parameter in this model is the interaction threshold, or bound of confidence: agents interact if their difference in opinions is smaller than some fixed value. As a result of their interaction, the opinions of the agents become closer by an amount proportional to their initial difference. 

There has been interest in going beyond this option to include the possibility of growing apart as a result of interaction, by including \emph{repulsive links} according to some rule. Jager \emph{et al.} \cite{Jager} interpret the bound of confidence as a latitude of acceptance of opinions, implying a willingness to move closer to those with whom we have some affinity  and, accordingly, they define a latitude of rejection. Drawing upon a social judgement theory, they assume that opinions get farther apart if their difference is greater than a given threshold. A somewhat similar reasoning was behind some recent modifications of opinion models to incorporate repulsive links \cite{mussot, sal}. However, the confidence bound can also be interpreted as simply a threshold for interaction, with no further implications on its outcome. Huet \emph{et al.} \cite{Huet} apply the concepts of dissonance theory by considering two dimensions. The rejection/attraction disposition on one dimen
 sion is conditioned by the disagreement/agreement in another dimension. Regardless of their repulsive or attractive disposition, agents only interact if their opinions are close enough, the reasoning being that if an agent has an \emph{a priori} rejection feeling towards someone, they feel uncomfortable if their positions are too close. 

Still, the reasons for rejection being the outcome of interaction are many, and do not confine themselves to the opinions in confrontation, in one or another dimension. Rejection can result from a rational discussion, when people realise that even though they share the same opinion, they do it for contradictory reasons, or from the desire to distinguish oneself from some individuals, to define a social status. In the present study, we model those many possible reasons as random, without considering another dimension. 

The previous considerations concern interactions between agents. But in a real society opinion evolution is also affected by external factors, like political propaganda or advertising. Carletti \emph{et al.} \cite{Carl} study the conditions for an efficient spreading of propaganda in the Deffuant \emph{et al.} model, and find that when the interaction threshold is small, propaganda can only have local effects. In this work, we show this is not the case under the presence of repulsive links: when agents prefer to have different opinions than some of their neighbours, consensus can be built around an external message, even in close-minded societies. 

This counterintuitive result is reminiscent of studies in which the presence of some kind of disorder - like noise \cite{HM,SRopi, SRopi2}, diversity \cite{TMTG06,DIRopi} or competitive interactions \cite{dac, fi4} - enhances the response to a weak time-dependant signal. Typically in those systems, the individual units have some stable states that can only be left or accessed by overcoming a barrier that a weak signal alone is not enough to surpass. Disorder cooperates with the signal by inducing collective switches at the signal rhythm between the stable states.

The novelty of the present study lies on the identification of a comparable phenomenon - an optimal response to a message resulting from the presence of disorder - in a system that does not have the usual ingredients. An agent can adopt any opinion on an interval, not having an intrinsic preferred state. But as a result of the collective dynamics, opinions can be fragmented into several non-interacting groups \cite{def}, some of which will be beyond the message's threshold of interaction. As we will see, it is the presence of repulsive links that enables agents to reach the basin of interaction of the signal. 

In the rest of the paper, after defining the model and the corresponding parameters, we present and analyse the results and summarize our main conclusions.

\section{Model}
\label{model}

We consider that an agent $i$, taken from a set of $i=1,\dots,N$ agents, holds at time $t$ an opinion $x_t^i$ expressing on a numerical scale his degree of agreement on a particular topic. The opinions take values on the interval $[0,1]$: values close to $0$ indicate a large degree of disagreement, and values close to $1$ a large degree of agreement with the topic in question. At time $t=0$, the opinions are independently drawn from a uniform random distribution in the interval $[0,1]$. We assign a weight $\omega_{ij}$ to the link connecting agents $i$ and $j$ and consider the symmetric case $\omega_{ij}=\omega_{ji}$ (or an undirected network). The weights take randomly positive or negative values: $\omega_{ij}=1$ with probability $(1-p)$ or $\omega_{ij}=-1$, with probability $p$. 

 The model is based on the rules introduced by Deffuant {\sl et al.}: At time $t$ two individuals, say $i$ and $j$, are randomly chosen. If their opinions are closer than the bound of confidence $\epsilon$, $|x_t^i-x_t^j|<\epsilon$, they converge or diverge as

 \begin{equation} \label{eq:rule}
 x_{t+\tau}^{i(j)} = x_{t}^{i(j)}+\mu \omega_{ij}(x_{t}^{j(i)}-x_{t}^{i(j)}),
 \end{equation} 
where the parameter $\mu$ mainly determines the speed of convergence or divergence. We will adopt from now on the value $\mu=0.5$, which minimises the transients. Note that as a consequence of repulsive interactions, this evolution rule could allow opinions to leave the interval $[0,1]$. To keep $x_t^i$ in the interval $[0,1]$, we impose the extra adsorbing boundary conditions: if $x_{t+\tau}^i<0$, $x_{t+\tau}^i=0$, and if $x_{t+\tau}^i>1$, $x_{t+\tau}^i=1$. By taking $\tau=1/N$, we define the usual unit of Monte Carlo time as $N$ updates.

To model the effect of advertising we add the rule that every $T/\tau$ agent-agent interactions, the entire population interacts simultaneously with an external message, or signal, $S$. That is, for every individual $i$, if $|x_t^i-S|<\epsilon$,
\begin{equation} 
\label{signal} 
x_{t+\tau}^{i} = x_{t}^{i}+\mu(S-x_{t}^{i}),
\end{equation} 
 where $S$ is a constant in the interval $[0,1]$.

Note that the original Deffuant $\emph{et al.}$ model \cite{def} with propaganda \cite{Carl} is recovered when $p=0$. 

\section{Results}
\label{results}

\begin{figure}
\center{
\includegraphics[width=6cm]{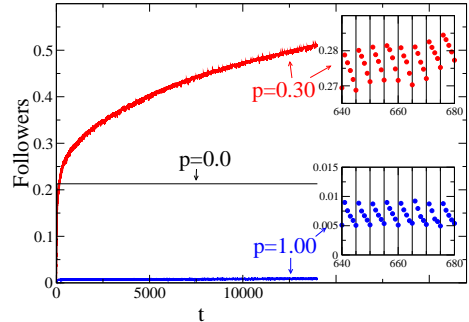}
\caption{(Colour online) Time evolution of the fraction of followers. $\epsilon=0.1$, $S=0.1$, $N=10^3$, $T=5$. In the insets, we have a closer view of the $p=0.3$ and $p=1.0$ cases. The vertical lines show the times where the signal was acting, and hence the number of followers increases. Averages over 100 runs. \label{pic}}}
\end{figure}

Figure \ref{pic} shows the simulations results for the number of agents whose opinion coincides with the propaganda - the ``followers''. We start our analysis by noting that the time evolution proceeds much slower when there is a combination of positive and repulsive links, compare $p=0.30$ with $p=0$ and $p=1$. We would reach asymptotically a steady state in the limit $t\to\infty$. However, we have decided not to focus on those asymptotic values since we argue that they can have no practical interest, as no social interaction can persist for an infinite time. Instead, all time average results of this paper refer to averages in the latest 20\% of the time, or $t\in [11200, 14000]$, an arbitrary choice, but we note that the qualitative results do not depend on the chosen time interval.

\begin{figure}
\center{
\includegraphics[width=7.5cm]{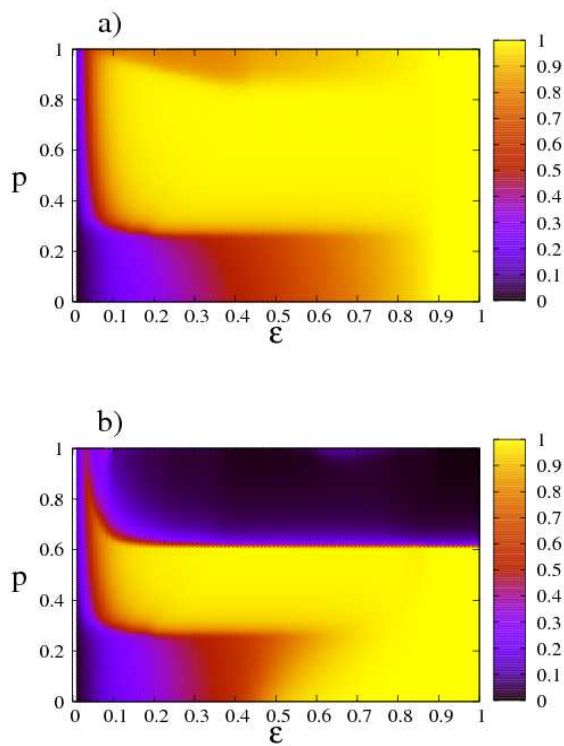}
\includegraphics[width=7.5cm]{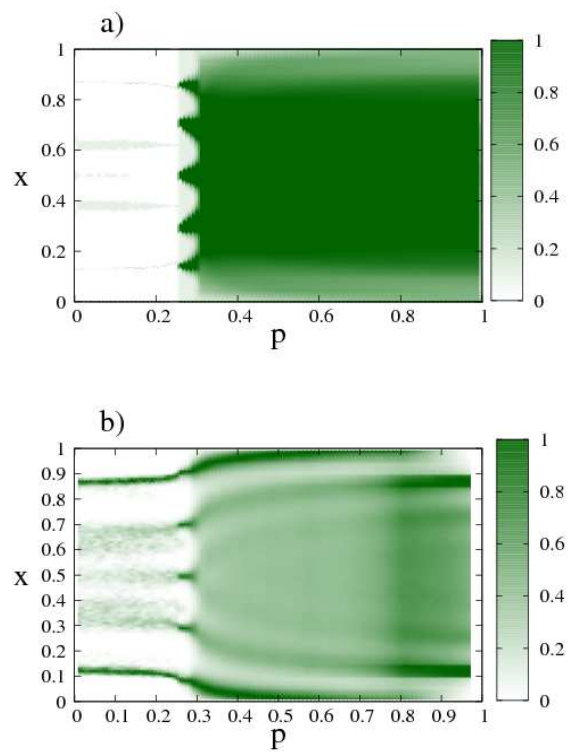}
\caption{(Colour online) Density plot of the fraction of followers as a function of the probability of repulsive links $p$ and the bound of confidence $\epsilon$ for messages of a) high frequency ($T=\tau=0.001$) and b) low frequency ($T=1$). The plots are the results of averages over $100$ runs in the case $S=0.1$ (extreme message) and $N=10^3$. Similar results are obtained for other values of $S$ and $N$, namely for $S=0.5$ (moderate message) for both $N=200$ and $N=10^3$.
\label{fol}}}
\end{figure}

\begin{figure}
\center{
\includegraphics[width=3.5cm, height=5.0cm, clip=true, trim=4 4 20 4]{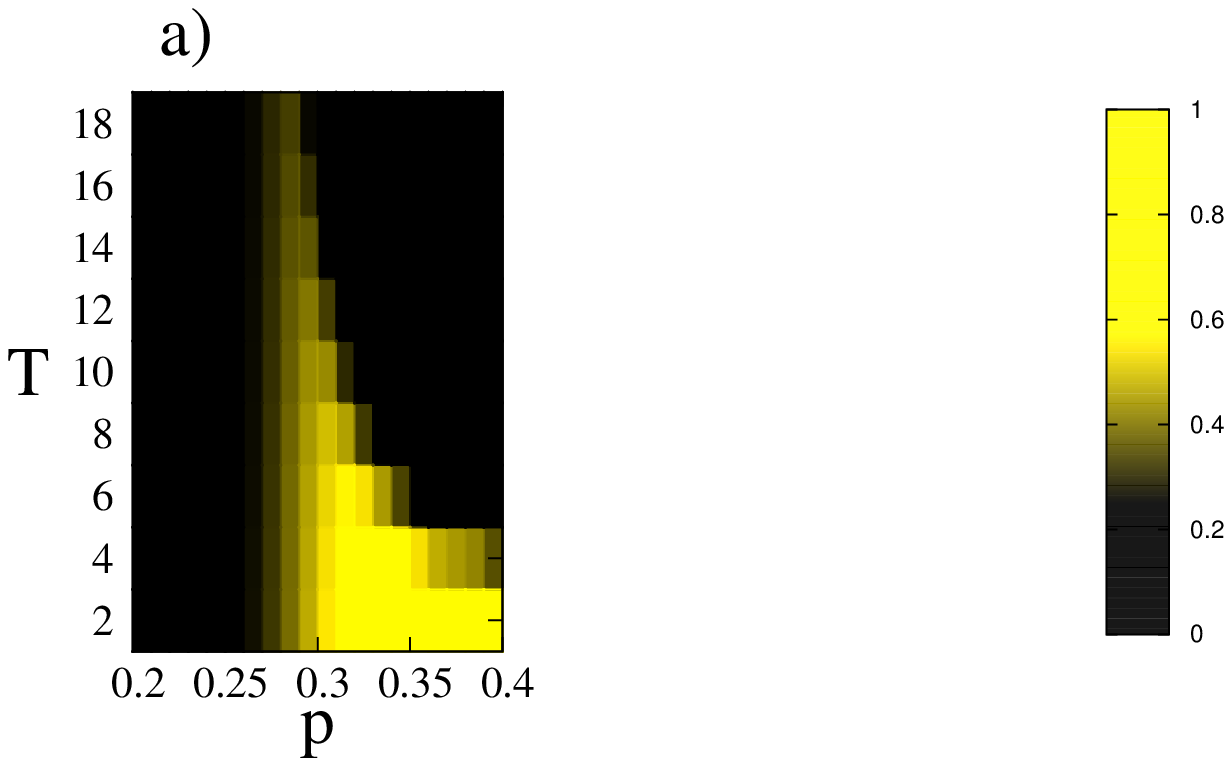}
\includegraphics[width=3.2cm, height=5.0cm, clip=true, trim=4 4 20 4]{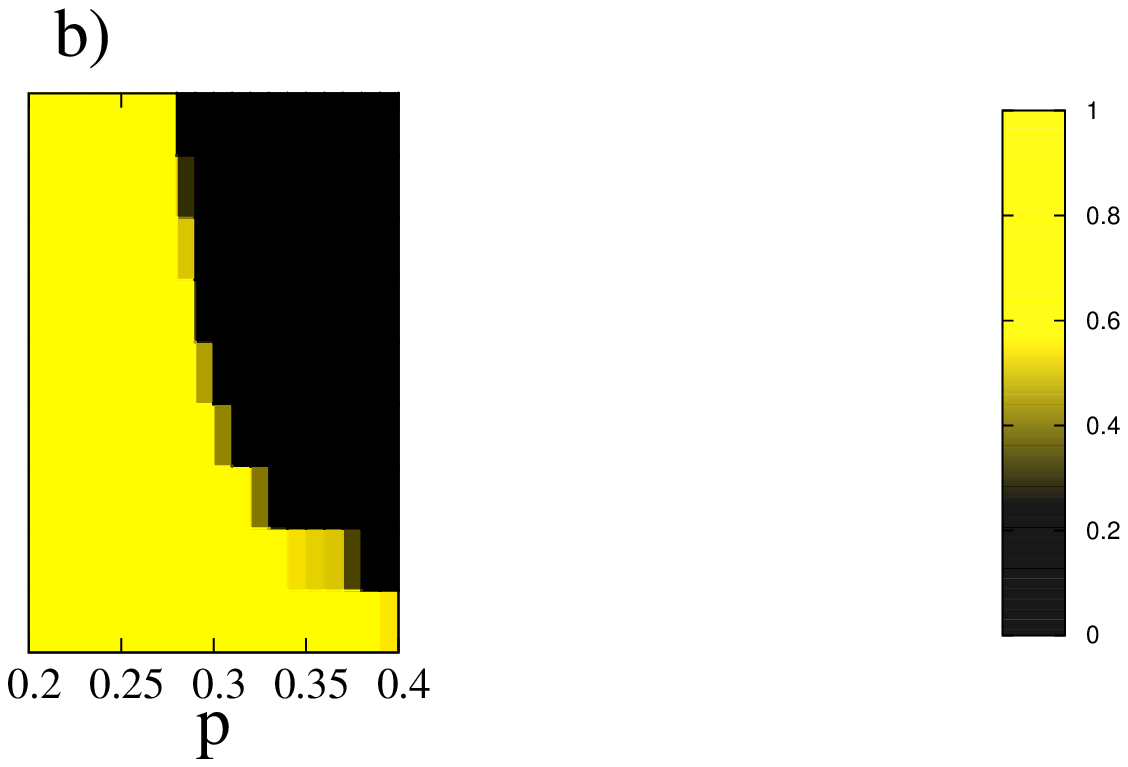}
\includegraphics[width=1.2cm, height=5.0cm, clip=true, trim=4 4 0 4]{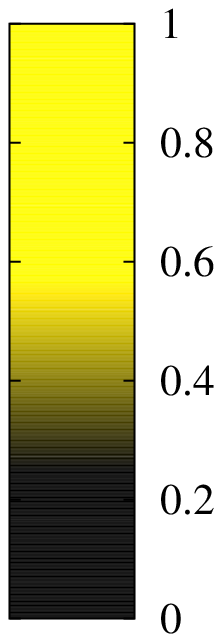}
\caption{(Colour online) The fraction of followers as a function of $p$ and $T$. As the period $T$ increases, it becomes increasingly harder for systems with a high probability of repulsive links to follow the message.  Parameters: a) $\epsilon=0.1$ and b) $\epsilon=0.7$. Other parameters: $S=0.1$, $N=10^3$. Averages over 100 runs.
\label{period}}}
\end{figure}

In fig. \ref{fol} we plot the average fraction of followers, as a function of the probability of the repulsive links $p$ and the bound of confidence $\epsilon$. For a given $p$, the adoption of the external message depends on the period $T$ of the signal, and on the the bound of confidence $\epsilon$. 
As noticed in \cite{Carl} for the $p=0$ case, when $\epsilon$ is sufficiently large, namely for $\epsilon>0.28$, the message can spread to all agents. In contrast, when all interactions are attractive and $\epsilon$ is small - a ``close-minded society'', where agents only interact with others whose opinion is very close - the message cannot convince the entire population \cite{Carl}. For $\epsilon<0.28$ and in the absence of advertising, opinions are fragmented into several major groups \cite{Lorenz} that don't interact with each other. Since some of the clusters will be outside the propaganda basin of attraction, the external message can only have local effects \cite{Carl}. In this region, the presence of a fraction of repulsive links is crucial for the message to spread to the entire population, as seen in fig. \ref{fol}. We also observe that a low-frequency signal (fig. \ref{fol}, b)) can not convince the entire population when there is a significant fraction of repulsive 
 links, a point to which we will return later in the paper (fig. \ref{period}). 

\begin{figure}
\center{
\includegraphics[width=7.5cm]{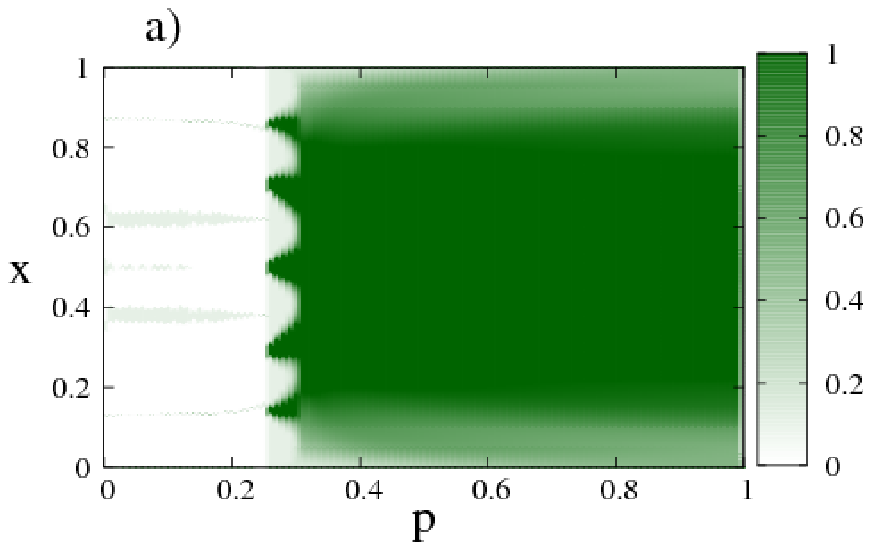}
\includegraphics[width=7.5cm]{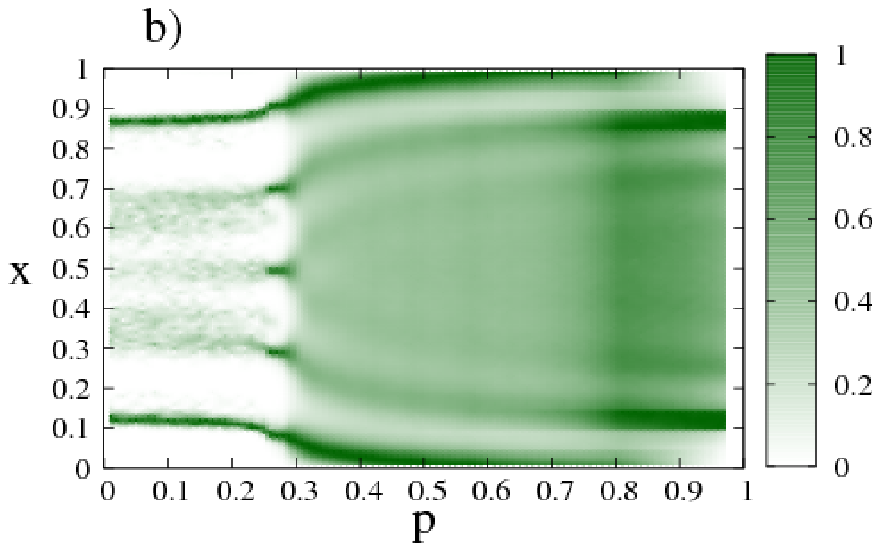}
\caption{(Colour online) Density plot of the distribution $P(x)$ of agents as a function of $p$ in the propaganda-free system for $\epsilon=0.1$ revealing a order-disorder transition at $p\approx0.30$. a) steady state results of the master equation. b) simulations with $N=10^4$ and averaged over $100$ runs, for $t=14000$. For better viewing, in the simulation results, the image was rescaled by dividing by the highest value at each probability and, in the master equation results, we considered each value greater than $1$ as $1$, and each value smaller than $0.1$ as $0.1$. 
\label{mese}}}
\end{figure}

Interestingly, the fraction of repulsive links where consensus around the signal starts to build, $p\approx0.30$, is remarkably independent on the specific characteristics of the message, like its frequency and value. This suggests that we must explore the effect repulsive interactions have on the system in the absence of signal, to find out what changes at that probability that can help the adoption of any kind of propaganda. 
In fig. \ref{mese} we present the results for the probability density function, $P(x)$, of the agents opinions coming from the numerical integration of the master equation \cite{ben, Miguel}, and from Monte Carlo simulations of the propaganda-free system. The master equation can be seen as the $N\to \infty$ limit of the Monte Carlo simulations approach: since those are done for finite $N$, there are some small differences in the results coming from the two methods. At $p=0$, clusters evolve to consensus regions, whereas at $p\approx0.30$ a distinctive pattern emerges in the distribution of opinions. Even though we can still distinguish clearly several peaks corresponding to higher concentrations of agents, the consensus inside a group is lost, and its boundaries are permeable, falling under the basin of interaction of a neighbouring one, something which turns out to be essential for optimal message reception, since it will allow the entire population to interact with an exter
 nal message.

However, the exposure to the message is only a precondition for following it, a second requirement being to stabilise agents in its position. Agents cease to negotiate when they share the same opinion, regardless of whether their connection is repulsive. In case propaganda convinces agents to adopt its message, the dynamics between those agents stops.

 The number of followers can only increase if propaganda convinces agents at a faster rate than the rate they disperse in its absence (insets fig. \ref{pic}). When the fraction of repulsive links becomes too high to stabilise a low-frequency signal, the number of followers decreases, as we see in figs. \ref{fol} and \ref{period}. If the bound of confidence $\epsilon$ is large, and the signal frequency is low, it is easier to form a consensus around the signal when there are few or none repulsive interactions (fig. \ref{fol}, b), and fig. \ref{period}, b)). When the signal frequency is low and $\epsilon$ is small, the region of optimal reception of the message shrinks until it coincides with the order-disorder transition (fig. \ref{fol}, b), and fig. \ref{period}, a)), where clusters are no longer consensus regions - facilitating interaction with neighbouring ones- but can still be clearly identified - agents don't spread much, which facilitates stabilisation. 

\begin{figure}
\center{
\includegraphics[width=4.2cm, height=4.0cm, clip=true, trim=5 2 30 15]{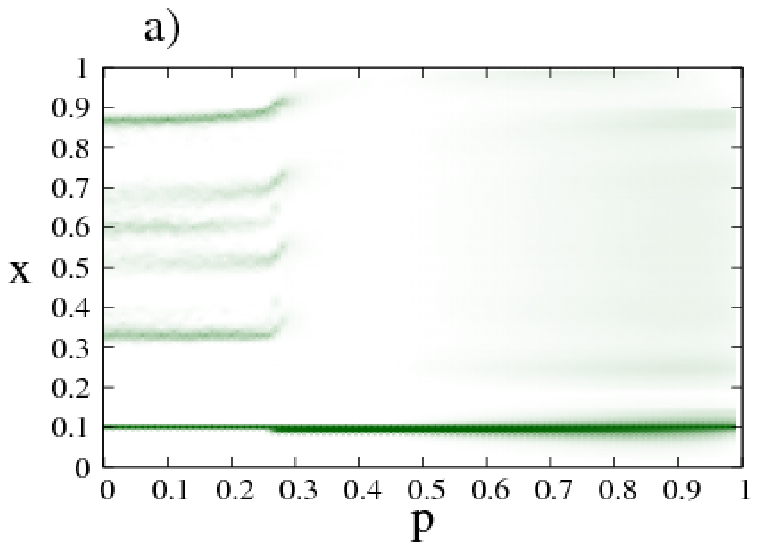}
\includegraphics[width=4.0cm, height=4.0cm, clip=true, trim=32 2 40 15]{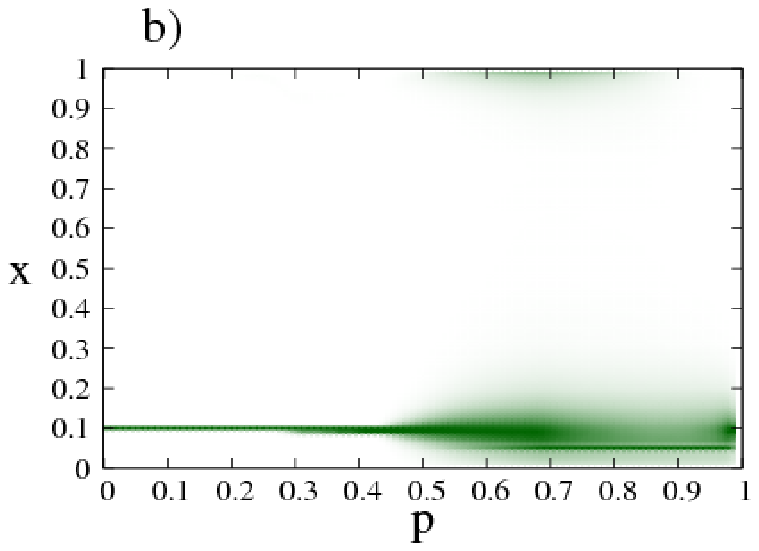}
\includegraphics[width=4.5cm, height=0.8cm, clip=true, trim=4 4 4 2]{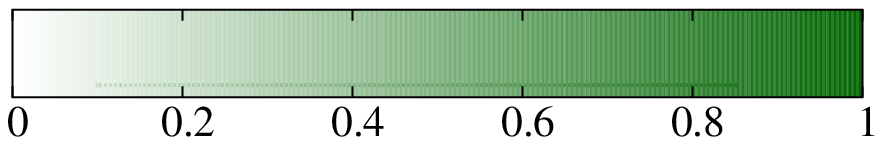}
\caption{(Colour online) Two-dimensional plot of the distribution of agents in opinion space, as a function of $p$ and with $S = 0.1$. Parameters: a) $\epsilon=0.1$ and b) $\epsilon=0.7$. For better viewing the image was rescaled by dividing by the highest value at each probability. Averages over 100 runs, $N=10^4$, $T=2$ and $S=0.1$.
\label{cluster}}}
\end{figure}

Figure \ref{cluster} shows how agents distribute themselves in the opinion space, as a function of $p$, when $S=0.1$ and $T=2$. We observe that when $\epsilon$ is small, (fig. \ref{cluster}, a)) and the value of $p$ is below the order-disorder transition region, the lack of followers reflects the formation of some groups that adopt an opinion outside the message's basin of attraction, and implies the permanent rejection of the message by a significant fraction of the population. As previously mentioned, when $\epsilon$ is large (fig. \ref{cluster}, b)) and all interactions are actractive, the message is adopted by the entire population.

\begin{figure}
\center{
\includegraphics[width=7.0cm]{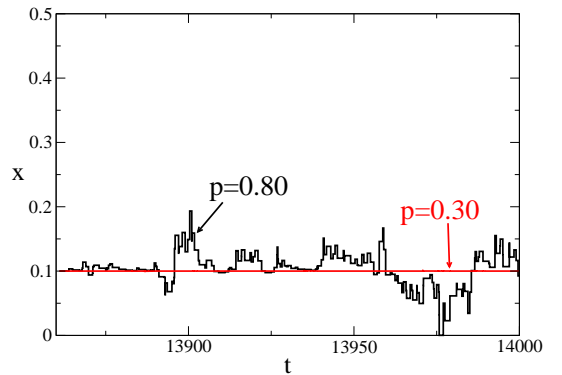}
\caption{(Colour online) Typical trajectory for one agent under signal influence with $\epsilon=0.1$ and other parameters as in fig. \ref{cluster}.
\label{traj}}}
\end{figure}

By contrast, when $p$ is too high to stabilize the signal, we don't observe the appearance of a plural society with well defined groups. Instead, most agents can spend some time adopting the message, forming a cluster around the propaganda, (fig. \ref{cluster}) that has its support base being continuously refreshed. In fig. \ref{traj} we plot a typical trajectory for one agent: when the probability of repulsive links is high, the agent changes its opinion constantly, without going far from the propaganda position. This represents a situation where a new level of opinion has been reached by most members of society, and yet there is still negotiation around details. Meanwhile, the situation of the few agents that don't gather around the external message depends on the type of society, as defined by its interaction threshold: for small $\epsilon$ (fig. \ref{cluster}, a)), they constantly change their opinion over all the opinion range, and are not beyond the possibility of still
  being convinced by the propaganda; while for large $\epsilon$ (fig. \ref{cluster}, b)) agents form a cluster in the extreme opposite position, if the external message was also extreme.

\section{Summary and Conclusions}
\label{conclusion}

In this work we have analysed the response to an external message, of a social system represented by the Deffuant \emph{et al.} model with a combination of positive and repulsive interactions. We focused in more detail on the case of a low confidence bound, or close-minded society, where the presence of a given fraction of repulsive links is required for the entire population to adopt the message. 

In that small $\epsilon$ case, the region of optimal response to the signal starts to build around an order-disorder transition region, that we identified resorting to the master equation and simulation results of the propaganda-free system. When the signal frequency decreases, this optimal region shrinks until it coincides with the order-disorder transition point. The same coincidence was noticed in previous studies \cite{TMTG06, dac, fi4} showing an optimal response induced by diversity or competitive interactions, and we conclude that it is expected in extended systems where some of the units can be in a state that is inaccessible to the external signal, whether it is because of the existence of a potential barrier like in the previous works, or because of a threshold of interaction. We saw that the presence of repulsive links and consequent dispersion of opinions is a necessary condition for a collective adoption of the message, by increasing the number of agents within t
 he reach of propaganda. Below the transition region, we assist to the formation of a plural society, because a substantial fraction of agents doesn't interact with propaganda. 

We discussed results concerning an intermittent propaganda that has always the same value $S$. We also tested the case of a sinusoidal propaganda, and found the same enhancement of the response for a certain fraction of repulsive links, that, not surprisingly, was harder to stabilise in the exact signal position. Since, unlike in previous studies \cite{TMTG06, dac, fi4}, agents don't have preferred opinions between which they can oscillate, the time-varying propaganda case corresponds simply to a very low frequency intermittent message, that as we saw can only receive a significant response in the order-disorder transition region. 

It is not a surprise that a close-minded society with strict agreement rules, or the paradigm of a very conservative society, is not open to outside influences. What is not so expected is to find out that the presence of repulsive links can in fact drive the population to form a consensus around an external message, regardless of whether the message is extreme or moderate. In this situation, and as a result of wanting to be apart, agents end up together sharing the same opinion. 

In this work we stress the importance that repulsive links have in the dynamics the Deffuant \emph{et al.} model. Further studies should address the effect that repulsive interactions have in the dynamics of other continuous opinion models with different interacting rules.

\acknowledgements
 We acknowledge financial support by the MEC (Spain) and FEDER (EU) through project FIS2007-60327. TVM acknowledges the support of FCT (Portugal) through Grant No.~SFRH/BD/23709/2005, MP is supported by the Belgian Federal Government (IAP project ''NOSY: Nonlinear systems, stochastic processes and statistical mechanics''). 

%
%
\bibliographystyle{eplbib}

\end{document}